\documentclass[aps,prb,showpacs,twocolumn]{revtex4}
\usepackage{graphicx}
\begin{document}
\title{Radiation-intensity and temperature dependence of microwave-induced 
magnetoresistance oscillations in high-mobility two-dimensional electron systems}
\author{X.L. Lei}
\affiliation{Department of Physics, Shanghai Jiaotong University,
1954 Huashan Road, Shanghai 200030, China}

\begin{abstract}
We present a detailed theoretical investigation on the radiation 
induced giant magnetoresistance oscillations recently discovered 
in high-mobility two-dimensional electron gas. Electron interactions 
with impurities, transverse and longitudinal acoustic phonons 
in GaAs-based heterosystems are considered simultaneously. 
Multiphoton-assisted impurity scatterings are shown to be the 
primary origin of the resistance oscillation. Based on the 
balance-equation theory developed for magnetotransport in Faraday 
geometry, we are able not only to reproduce the observed period, 
phase and the negative resistivity of the main oscillations, but also 
to predict the secondary peak/valley structures relating to two-photon 
and three-photon processes. The dependence of the magnetoresistance 
oscillation on microwave intensity, the role of dc bias current 
and the effect of elevated electron temperature are discussed.  
Furthermore, we propose that the temperature-dependence of the resistance 
oscillation stems from the growth of the Landau level broadening due to 
the enhancement of acoustic phonon scattering with increasing lattice 
temperature. The calculated temperature-variation of the oscillation 
agrees well with experimental observations.
\end{abstract}


\maketitle

\section{Introduction}
Tremendous interest in magneto-transport in two-dimensional electron 
system (2DES) 
has recently revived since the experimental discovery of giant oscillations 
of the longitudinal resistance as a function of the magnetic field 
in high mobility two-dimensional (2D) electron gas (EG) subjected to 
a microwave radiation,\cite{Zud01,Ye,Mani05507} particularly following
the recent observations of "zero-resistance" states in very clean samples 
by two independent groups.\cite{Mani,Zud03,Mani06388,Zud06508}
These radiation-induced oscillations of the longitudinal magnetoresistivity 
$R_{xx}$ are periodical in inverse magnetic field $1/B$ but, unlike the 
well-known Shubnikov-de Haas (SdH) oscillation whose period is controlled 
by the electron density $N_{\rm e}$,\cite{Kohn} their periods 
are determined by the radiation frequency $\omega$.
The observed $R_{xx}$ oscillations exhibit a smooth magnetic-field variation 
with the resistivity maxima at $\omega/\omega_c=j-\delta_{-}$ 
and minima at $\omega/\omega_c=j+\delta_{+}$ ($\omega_c$ is the cyclotron
frequency, $j=1,2,3...$) having positive $\delta_{\pm}$ 
ranging around $0.1-0.25$.\cite{Mani,Zud06508} 
The resistivity minimum goes downward 
with increasing sample mobility and/or increasing radiation intensity 
until a vanishing resistance state shows up, while the Hall resistivity 
keeps the classical form $R_{xy}=B/e N_{\rm e}$ 
with no sign of quantum Hall plateau 
over the whole magnetic field range exhibiting $R_{xx}$ oscillation.
Later independent experiments\cite{Dor,Will} confirmed these results
and the corresponding zero-conductance states were also observed 
in the Corbino samples.\cite{Yang}

To explore the origin of these peculiar "zero-resistance"
states, different mechanisms have been 
suggested.\cite{Durst,Andreev,Anderson,Xie,Phil,Koul} 
As is shown by Andreev {\it et al.}\cite{Andreev} that a negative linear 
conductance implies that the zero current state is intrinsically unstable:
the system spontaneously develops a non-vanishing local current density
which is determined by the condition that the component of electric field
parallel to the local current vanishes. Thus the appearance of negative 
longitudinal resistivity or conductivity in a uniform model suffices 
to explain the observed vanishing resistance. 
The possibility of absolute negative photoconductance in a 2DES subject to 
a perpendicular magnetic field was first explored 30 years ago by 
Ryzhii.\cite{Ryz,Ryz86} Experimentally Keay {\it et al}\,\cite{Keay}reported 
the observation of absolute negative conductance in sequential resonant tunneling
superlattices driven by intense terahertz radiation. 
Recent works\cite{Durst,Anderson,Xie} indicated that the periodical 
structure of the density of states (DOS) of the 2DEG in a magnetic field 
and the photon-excited electron scatterings by impurities are the main
origin of the magnetoresistance oscillations. 
Durst {\it et. al.}\cite{Durst} proposed a microscopic analysis 
for the conductivity assuming a $\delta$-correlated disorder and a simple  
form of the 2D electron self-energy oscillatory with the magnetic field, 
obtaining the correct period, phase and the possible negative resistivity. 
Shi and Xie\cite{Xie} gave a similar result using the Tien and Gorden
current formula\cite{Tien} for photon-assisted coherent tunneling.
A more quantitative theoretical description was reported recently using  
a balance equation approach developed for radiation-induced magnetotransport 
in Faraday geometry,\cite{Lei04687} not only reproducing the correct period, 
phase and the negative resistivity of the main oscillations, but also 
predicting the secondary peaks and additional maxima and minima observed 
in the experiment,\cite{Mani,Zud03,Zud06508,Dor} and identifying them 
as arising from double- and triple-photon processes. 
A quantum Boltzmann equation approach based on self-consistent Born 
approximation for large filling factors has also been 
presented very recently, taking account of the elastic (impurity) 
scattering as the major mechanism
for radiation-induced magnetoresistance oscillation.\cite{Vav05478}
In addition to photon-assisted impurity scattering referred above as the 
mechanism of the absolute negative conductivity (ANC), other possible
mechanisms were also explored in the literature. Ryzhii {\it et al.} proposed that
acoustic phonon scattering\cite{Ryz05199,Ryz05454} 
and heating of electrons\cite{Ryz05484} could also serve as the mechanisms of ANC
in 2DES, and made a further attempt to connect them with experiments.\cite{Ryz07223}
 
One of the most important and interesting features of this phenomenon is 
its sensitive temperature dependence. The "zero-resistance"
states and radiation-induced magnetoresistance oscillations show up
strongly only at low temperatures typically around $T=1$\,K or lower. 
At fixed microwave power with increasing temperature, not only the zero-resistance 
regions become narrower and eventually disappear, the whole oscillatory 
structure (peaks and valleys) diminish as well. At temperature $T\geq$\,4-5\,K,
oscillatory structure disappears completely and the resistivivty 
$R_{xx}$-versus-magnetic field becomes essentially structureless.\cite{Mani,Zud03} 
The temperature-variation of $R_{xx}$ at deepest minima exhibits approximate
activated-type behavior $\exp(T_0/T)$. However, the activation energies
observed by both groups are very high: up to 10\,K and 20\,K at $j=1$ minimum
respectively.\cite{Mani,Zud03} These values are about an order of magnitude 
higher than the microwave photon energy ($\omega\sim $3-5\,K)
and Landau-level spacing ($\omega_c\leq$2\,K). Furthermore, different $T_0$ values
observed by the two groups indicate that the disappearing speed of the 
oscillatory structure with increasing temperature 
is sample dependent.\cite{Mani,Zud03}
To my knowledge, there has been no theoretical attempt 
to explain this temperature dependence, 
except that a conjecture of the formation of an energy gap around
the Fermi surface is suggested under microwave irradiation around the resistance
minima.\cite{Mani}
 
The recently constructed balance-equation model,\cite{Lei04687,Liu03} which
provides a quantitative and tractable approach to radiation-induced 
magnetotransport in Faraday geometry, enables us not only to analyze
the magnetoresistance oscillation, its dependence on the radiation
intensity, but also to deal with its temperature variation 
in a comprehensive way. We suggest that the temperature suppression
of the magnetoresistance oscillation in these high-mobility 2DES  
comes mainly from the growth of the Landau level broadening due to the 
rapid enhancement of acoustic phonon scatterings with increasing temperature
at this low temperature range. 

In this paper we will carry out a detailed theoretical investigation on the 
different aspects of radiation-induced magnetoresitance oscillation. 
The paper is organized as follows. For the convenience and completeness 
we present the general theoretical model and formulation in Section II.
As a typical example we analyze GaAs-based systems for which the relevant material
parameters are discussed in the Section III. Section IV concentrates 
on the transport properties of GaAs-based heterosystems at lattice temperature 
$T=1$\,K. 
We will give a detailed discussion on the impurity and acoustic phonon scattering
related linear and nonlinear magnetoresistance induced by the irradiation of 0.1\,THz
microwave of different intensities, and the effect of elevated electron temperature.   
Section V is devoted to the analyses on the lattice-temperature dependence 
of the magnetoresistance oscillation. Finally, a brief summary is given in Section VI.

\section{formulation}
\subsection{Balance equations in crossed electric and magnetic fields}
     
The experiments allow us to assume the 2DEG being in extended states over 
the magnetic field range relevant to this phenomenon.
For a general treatment,
we consider $N_{\rm e}$ electrons in a unit area of a quasi-2D system
in the $x$-$y$ plane with a confining potential $V(z)$ in the $z$-direction. 
These electrons, besides interacting with each other, are scattered by
random impurities/disorders and by phonons in the lattice. 
To include possible elliptically polarized microwave illumination we assume that 
a uniform dc electric field ${\bf E}_0$ and a high-frequency (HF) ac field 
of frequency $\omega$,
\begin{equation}
{\bf E}_t\equiv{\bf E}_s \sin(\omega t)+{\bf E}_c\cos(\omega t),\label{Et}
\end{equation} 
 are applied in the $x$-$y$ plane, 
together with a magnetic field ${\bf B}=(0,0,B)$ along the $z$ direction.
In terms of the 2D center-of-mass momentum and coordinate of
the electron system,\cite{Ting,Lei85,Lei851} which are defined as 
${\bf P}\equiv\sum_j {\bf p}_{j\|}$ 
and ${\bf R}\equiv N_{\rm e}^{-1}\sum_j {\bf r}_{j\|}$  
with ${\bf p}_{j\|}\equiv(p_{jx},p_{jy})$ and ${\bf r}_{j\|}\equiv (x_j,y_j)$
being the momentum and coordinate of the $j$th electron in the 2D plane,
and the relative electron momentum and coordinate 
${\bf p}_{j\|}'\equiv{\bf p}_{j\|}-{\bf P}/N_{\rm e}$ and 
${\bf r}_{j\|}'\equiv{\bf r}_{j\|}-{\bf R}$,
the Hamiltonian of the system can be written as the sum of 
a center-of-mass part $H_{\rm cm}$
and a relative electron part $H_{\rm er}$ 
(${\bf A}({\bf r})$ is the vector potential of the ${\bf B}$ field),
\begin{eqnarray}
H_{\rm cm}=\frac 1{2N_{\rm e}m}({\bf P}-N_{\rm e}e{\bf A}({\bf
R}))^2-N_{\rm e}e({\bf E}_{0}+{\bf E}_t)\cdot {\bf R},&&\label{Hcm}\\
H_{\rm er}=\sum_{j}\Big[\frac{1}{2m}\left({\bf p}_{j\|}'-e{\bf A}
({\bf r}_{j\|}')\right)^{2}
+\frac{p_{jz}^2}{2m_z}+V(z_j)\Big]\,\,&&\nonumber\\
+\sum_{i<j}V_c({\bf r}_{i\|}'-{\bf r}_{j\|}',z_i,z_j),\,\,\,\,\,\,&&\label{Her}
\end{eqnarray}
together with electron-impurity and electron-phonon interactions $H_{\rm ei}$ 
and $H_{\rm ep}$. Here $m$ and $m_z$ are respectively the electron effective mass
parallel and perpendicular to the plane, and $V_c$ stands for the electron-electron
Coulomb interaction. 
It should be noted that the uniform electric field (dc and ac) appears only in 
$H_{\rm cm}$, and that $H_{\rm er}$ is just the Hamiltonian of a quasi-2D system 
subjected to a magnetic field.
The coupling between the center-of-mass and the relative electrons exists via 
the electron-impurity and electron-phonon interactions. Our treatment starts with
the Heisenberg operator equations for the rates of changes of the center-of-mass 
velocity 
$\dot{\bf V}=-i[{\bf V},H]+\partial{\bf V}/\partial t$, with ${\bf V}=-i[{\bf R},H]$,
and of the relative electron energy 
$\dot{H}_{\rm er}=-{\rm i}[H_{\rm er},H]$, and proceeds with the determination of their
statistical averages.

As proposed in Ref.\,\onlinecite{Lei85}, the center-of-mass coordinate ${\bf R}$ 
and velocity ${\bf V}$ can be treated classically, i.e. as the time-dependent
expectation values of the center-of-mass coordinate and velocity,
${\bf R}(t)$ and ${\bf V}(t)$, such that ${\bf R}(t)-{\bf R}(t^{\prime})
=\int_{t^{\prime}}^t{\bf V}(s)ds$.
We are concerned with the steady transport state
under an irradiation of single frequency and focus on the 
photon-induced dc resistivity and the energy absorption of the HF field. 
These quantities are directly related to the time-averaged and/or base-frequency 
oscillating components of the center-of-mass velocity.
Although higher harmonics of the current 
may affect the dc and lower harmonic terms of the drift velocity 
through entering the damping force and energy exchange rates 
in the resulting equations, in an ordinary semiconductor the power of 
even the third harmonic current is rather weak as compared to the fundamental. 
For the HF field intensity in the experiments, 
the effect of higher harmonic current is safely negligible. 
Hence, it suffices to assume that the center-of-mass 
velocity, i.e. the electron drift velocity, consists of a dc
part ${\bf v}_0$ and a stationary time-dependent part ${\bf v}(t)$ of the form
\begin{equation}
{\bf V}(t)={\bf v}_0+{\bf v}_1 \cos(\omega t)+{\bf v}_2 \sin(\omega t).
\end{equation}
With this, the exponential factor in the operator equations can be expanded 
in terms of Bessel functions ${\rm J}_n(x)$:
\begin{eqnarray}
\hspace*{-1.3cm}&&{\rm e}^{-{\rm i}{\bf q}\cdot \int_{t^{\prime }}^{t}{\bf V}(s)ds}
=\sum_{n=-\infty }^{\infty }{\rm J}_{n}^{2}(\xi ){\rm e}^{{\rm i}({\bf q}\cdot 
{\bf v}_0-n\omega) (t-t^{\prime
})}+\nonumber\\
&&\sum_{m\neq 0}{\rm e}^{{\rm i}m(\omega t-\varphi )}\sum_{n=-\infty }^{\infty
}{\rm J}_{n}(\xi ){\rm J}_{n-m}(\xi ){\rm e}^{{\rm i}({\bf q}\cdot 
{\bf v}_0-n\omega) (t-t^{\prime })}.\nonumber
\end{eqnarray}
Here the argument in the Bessel functions 
\begin{equation}
\xi\equiv \frac{1}{\omega}\left[({\bf q}_\|\cdot {\bf v}_1)^2+
({\bf q}_\|\cdot {\bf v}_2)^2\right]^{\frac{1}{2}}
\end{equation}
and  
$\tan \varphi=({\bf q}\cdot {\bf v}_2)/({\bf q}\cdot {\bf v}_1)$.
 On the other hand, for 2D systems having 
electron sheet density of order of 10$^{15}$ m$^{-2}$, 
the intra-band and inter-band Coulomb interactions are sufficiently strong 
that it is adequate to describe the relative-electron transport state using a single 
electron temperature $T_{\rm e}$. Except this, the electron-electron interaction
is treated only in a mean-field level under random phase approximation 
(RPA).\cite{Lei85,Lei851} 
For the determination of unknown parameter ${\bf v}_0$, ${\bf v}_1$, ${\bf v}_2$, 
and ${T_{\rm e}}$, it suffices to know the damping force 
up to the base frequency oscillating term 
${\bf F}(t)= {\bf F}_0+{\bf F}_s\sin(\omega t)+{\bf F}_c\cos(\omega t)$, 
and the energy-related 
quantities up to the time-average term. We finally obtain the  
force and energy balance equations:
\begin{eqnarray}
0&=&N_{\rm e}e{\bf E}_{0}+N_{\rm e} e ({\bf v}_0 \times {\bf B})+
{\bf F}_0,\label{eqv0}\\
{\bf v}_{1}&=&\frac{e{\bf E}_s}{m\omega}+\frac{{\bf F}_s}{N_{\rm e}m\omega }
-\frac{e}{m\omega }({\bf v}_{2}\times
{\bf B}),\label{eqv1}\\
-{\bf v}_{2}&=&\frac{e{\bf E}_c}{m\omega}+\frac{{\bf F}_c}{N_{\rm e}m\omega }
-\frac{e}{m\omega }({\bf v}_{1}
\times {\bf B}),\label{eqv2}
\end{eqnarray}
\begin{equation}
N_{\rm e}e{\bf E}_0\cdot {\bf v}_0+S_{\rm p}- W=0.
\label{eqsw}
\end{equation}
Here
\begin{eqnarray}
{\bf F}_{0}=\sum_{{\bf q}_\|}\left| U({\bf q}_\|%
)\right| ^{2}%
\sum_{n=-\infty }^{\infty }{\bf q}_\|{\rm J}_{n}^{2}(\xi )\Pi _{2}({\bf %
q}_\|,\omega_0-n\omega )\,\,\,\,\,&&\nonumber\\
+\sum_{{\bf q}}\left| M({\bf q})\right|
^{2}\sum_{n=-\infty
}^{\infty }{\bf q}_\|{\rm J}_{n}^{2}(\xi )\Lambda _{2}({\bf q},\omega_0+\Omega _{{\bf 
q}}-n\omega ),&&
 \label{eqf0}
\end{eqnarray}
is the time-averaged damping force, $S_{\rm p}$ is the time-averaged rate of the 
electron energy-gain from the HF field,
$\frac{1}{2}N_{\rm e}e({\bf E}_s\cdot{\bf v}_2+{\bf E}_c\cdot{\bf v}_1)$,
which can be written in a form obtained 
from the right hand side of Eq.\,(\ref{eqf0}) by replacing the ${\bf q}_\|$ factor with 
$n \omega$, 
and $W$ is the time-averaged rate of the electron energy-loss due to coupling 
with phonons, whose expression can be obtained from the second term on the right hand 
side of Eq.\,(\ref{eqf0}) by replacing the ${\bf q}_\|$ factor 
with $\Omega_{\bf q}$, the energy of a wavevector-${\bf q}$ phonon. 
The oscillating frictional force amplitudes 
${\bf F}_s\equiv {\bf F}_{22}-{\bf F}_{11}$ and 
${\bf F}_c\equiv {\bf F}_{21}+{\bf F}_{12}$ are given by ($\mu=1,2$)
\begin{widetext}
\begin{eqnarray}
{\bf F}_{1\mu}=-\sum _{{\bf q}_\|}{\bf q}_\|\eta_{\mu}| U({\bf q}_\|%
)| ^{2}\sum_{n=-%
\infty }^{\infty }\left[ {\rm J}_{n}^{2}(\xi )\right] ^{\prime }\Pi _{1}(%
{\bf q}_\|,\omega_0-n\omega )
- 
\sum_{\bf q}{\bf q}_\|\eta_{\mu}| M({\bf q})|
^{2}\sum_{n=-\infty
}^{\infty }\left[ {\rm J}_{n}^{2}(\xi )\right] ^{\prime }\Lambda _{1}({\bf q%
}, \omega_0+\Omega _{{\bf q}}-n\omega ),\,\,\,\,&&\label{eqf1u}\\ 
{\bf F}_{2\mu}=\sum_{{\bf q}_\|}{\bf q}_\|\frac{\eta_{\mu}}
{\xi}| U({\bf q}_\|)| ^{2}%
\sum_{n=-\infty }^{\infty }2n{\rm J}_{n}^{2}(\xi )\Pi _{2}({\bf %
q}_\|,\omega_0-n\omega )
+ 
\sum_{{\bf q}}{\bf q}_\|\frac{\eta_{\mu}}{\xi}| M({\bf q})|^{2}\sum_{n=-\infty
}^{\infty }2n{\rm J}_{n}^{2}(\xi )\Lambda _{2}({\bf q},\omega_0+\Omega _{\bf q}-n\omega 
).
\,\,\,\,\,\,\,\,\,
&&\label{eqf2u}
 \end{eqnarray}
\end{widetext}
In these expressions,
$\eta_{\mu}\equiv {\bf q}_\|\cdot {\bf v}_{\mu}/\omega \xi$;
$\omega_0\equiv {\bf q}_\|\cdot {\bf v}_0$;
$U({\bf q}_\|)$ and $M({\bf q})$ stand for effective impurity and phonon
scattering potentials, 
$\Pi_2({\bf q}_\|,\Omega)$ and
$
\Lambda_2({\bf q},\Omega)=2\Pi_2({\bf q}_\|,\Omega)
[n(\Omega_{\bf q}/T)-n(\Omega/T_{\rm e})]
$\,(with $n(x)\equiv 1/({\rm e}^x-1)$)
are the imaginary parts of the electron density correlation function 
and electron-phonon correlation function in the presence of the magnetic field.
$\Pi_1({\bf q}_\|,\Omega)$ and  $\Lambda_1({\bf q},\Omega)$
are the real parts of these two correlation functions.

The HF field enters through the argument $\xi$ of the Bessel functions 
in ${\bf F}_0$, ${\bf F}_{\mu\nu}$, $W$ and 
$S_{\rm p}$. Compared with that without the HF field 
($n=0$ term only),\cite{Lei98} we see that in an electron gas 
having impurity and/or phonon scattering (otherwise homogeneous),
a HF field of frequency $\omega$ opens additional channels for electron 
transition: an electron in a state can absorb or emit one or several photons
and scattered to a differnt state with the help of impurities and/or phonons.
The sum over $|n|\geq 1$ represents contributions of single and multiple
photon processes of frequency-$\omega$ photons. 
These photon-assisted scatterings help to transfer
energy from the HF field to the electron system ($S_{\rm p}$) 
and give rise to additional damping force on the moving electrons.

Eqs.\,(\ref{eqv0})-(\ref{eqsw}) form a closed set of equations for the 
determination of parameters ${\bf v}_0$, ${\bf v}_1$ ${\bf v}_2$ and $T_{\rm e}$  
when ${\bf E}_0$, ${\bf E}_c$ and ${\bf E}_s$ are given in a 2D system subjected 
to a magnetic field $B$ at temperature $T$. 
Thus they provide 
a comprehensive and quantitative description of transport and optical properties 
of magnetically-biased quasi-2D semiconductors subjected to a dc bias and a HF 
radiation field in Faraday geometry.

Note that ${\bf v_1}$ and ${\bf v}_2$ 
always exhibit cyclotroresonance in the range $\omega\sim\omega_c$, 
as can be seen from Eqs.(\ref{eqv1}) and (\ref{eqv2}) rewritten in the form
\begin{eqnarray}
&&{\bf v}_{1}=\frac{\omega^2}{(\omega^2-\omega_c^2)}
\left\{\frac{e}{m\omega}\left[{\bf E}_s+\frac{e}{m\omega }(
{\bf E}_c\times{\bf B})\right]\right.\nonumber\\
&&\hspace{1.0cm}+\left.\frac{1}{N_{\rm e}m\omega }
\left[{\bf F}_s+
\frac{e}{m\omega }(
{\bf F}_c\times
{\bf B})\right]\right\},\label{vv1}\\
&&{\bf v}_{2}=\frac{\omega^2}{(\omega_c^2-\omega^2)}
\left \{\frac{e}{m\omega}\left[{\bf E}_c-\frac{e}{m\omega }(
{\bf E}_s\times{\bf B})\right]\right.\nonumber\\
&&\hspace{1.0cm}+\left.\frac{1}{N_{\rm e}m\omega }
\left[{\bf F}_c-
\frac{e}{m\omega }
({\bf F}_s\times
{\bf B})\right]\right\}.\label{vv2}
\end{eqnarray}
Therefore, the argument $\xi$ may be significantly different from that of the 
corresponding Bessel functions in the case without a magnetic field or 
with a magnetic field in Voigt configuration, where the electron motion 
is not affected by the magnetic field.\cite{Lei98} 
On the other hand, impurity and phonon scatterings can affect $\xi$ through
the damping forces ${\bf F}_s$ and ${\bf F}_c$.
Eqs.\,(\ref{vv1}) and (\ref{vv2}) when neglecting the damping
forces ${\bf F}_s=0={\bf F}_c$, yield a Bessel-function argument 
$\xi$ equivalent to that used in the early literature\cite{Malov1973, Vyu1977}. 
The approximation of neglecting damping forces is valid only in weak scattering limit
and away from cyclotroresonance.  Depending on ${\bf v}_1$ and ${\bf v}_2$, 
the damping forces ${\bf F}_s$ and ${\bf F}_c$ in Eqs.\,(\ref{vv1}) and (\ref{vv2})
are important not only in the general scattering case over the whole magnetic-field 
range but also in weak scattering case in the vicinity of cyclotron resonance in that 
they remove the divergence and yield finite 
oscillation velocities ${\bf v}_1$ and ${\bf v}_2$ at $\omega=\omega_c$. 

\subsection{Longitudinal and transverse resistivities}
The nonlinear resistivity in the presence of a high-frequency
field is easily obtained from Eq.\,(\ref{eqv0}).
Taking ${\bf v}_0$ to be in the $x$ direction,
${\bf v}_0=(v_{0x},0,0)$, we immediately get the
transverse and longitudinal resistivities 
\begin{eqnarray}
R_{xy}&\equiv &\frac{E_{0y}}{N_{\rm e}ev_{0x}}=\frac{B}{N_{\rm e}e},\label{rxy}\\ 
R_{xx}&\equiv &\frac{E_{0x}}{N_{\rm e}ev_{0x}}
=-\frac{F_0}{N_{\rm e}^2 e^2 v_{0x}}.\label{rxx}
\end{eqnarray}
 We see that 
in the present model (with parabolic energy spectrum) the transverse resistivity 
$R_{xy}$ remains the classical form, with no change in the presence 
of HF radiation and/or dc current. 
The longitudinal resistivity $R_{xx}$, on the other hand, can be strongly affected 
by the irradiation also by dc bias. 
The linear longitudinal resistivity is 
the weak dc current limit ($v_{0x}\rightarrow 0$): 
\begin{eqnarray}
R_{xx}&=&-\sum_{{\bf q}_\|}q_x^2\frac{|
U({\bf q}_\|)| ^2}{N_{\rm e}^2 e^2}\sum_{n=-\infty }^\infty {\rm J}_n^2(\xi)\left. 
\frac {\partial \Pi_2}{\partial\, \Omega }\right|_{\Omega =n\omega }\nonumber\\
&&- \sum_{ {\bf q}} q_x^2\frac{\left| M ( {\bf
q})\right| ^2}{N_{\rm e}^2 e^2}\sum_{n=-\infty }^\infty {\rm J}_n^2(\xi)\left. 
\frac {\partial \Lambda_2}{\partial\, \Omega }\right|_{\Omega =\Omega_{{\bf 
q}}+n\omega}.
\,\,\,\,\,\,\,\label{lrxx}
\end{eqnarray}

Note that although according to Eqs.\,(\ref{eqf0}), (\ref{rxx}) and (\ref{lrxx}), the 
linear and nonlinear longitudinal magnetoresistivity $R_{xx}$ can formally written 
as the sum of contributions from various individual scattering mechanisms,
all the scattering mechanisms have to be taken into account simultaneously 
in solving the momentum- and energy-balance equations (\ref{eqv1}), (\ref{eqv2})
and (\ref{eqsw}) for ${\bf v}_1$, ${\bf v}_2$ and $T_{\rm e}$,
which enter the Bessel functions and other parts in the expression of $R_{xx}$.

\subsection{Landau-level broadening}

In the present model the effects of interparticle Coulomb interactions are 
included in 
the electron complex density correlation function $\Pi({\bf q}_{\|},\Omega)
=\Pi_1({\bf q}_{\|},\Omega)+i\Pi_2({\bf q}_{\|},\Omega)$,
which, in the random phase approximation, can be expressed as
\begin{equation}
\Pi({\bf q}_{\|},\Omega)=\frac{\Pi_0({\bf q}_{\|},\Omega)}{\epsilon({\bf 
q}_\|,\Omega)},
\end{equation}
where
\begin{equation}
\epsilon({\bf q}_\|,\Omega)\equiv 1-V(q_{\|})\Pi_0({\bf q}_{\|},\Omega)
\end{equation}
is the complex dynamical dielectric function, 
\begin{equation}
V(q_{\|})=\frac{e^2}{2\epsilon_0\kappa q_{\|}}H(q_{\|})
\end{equation}
is the effective Coulomb potential with $\kappa$ the dielectric constant of the 
material 
and $H(q_{\|})$ a 2D wavefunction-related overlapping integration,\cite{Lei851} 
$\Pi_0({\bf q}_{\|},\Omega)=\Pi_{01}({\bf q}_{\|},\Omega)+i\Pi_{02}({\bf 
q}_{\|},\Omega)$ 
is the complex density correlation function of the independent electron 
system in the presence
of the magnetic field. With this dynamically screened density correlation function 
the collective plasma modes of the 2DES are incorporated. Disregard these 
collective modes one can just use a static screening 
$\epsilon({\bf q}_{\|},0)$ instead. 

The $\Pi_{02}({\bf q}_{\|}, \Omega)$ function of a 2D
system in a magnetic field can be written in terms of Landau representation:\cite{Ting}
\begin{eqnarray}
&&\hspace{-0.7cm}\Pi _{02}({\bf q}_{\|},\Omega ) =  \frac 1{2\pi
l_{\rm B}^2}\sum_{n,n'}C_{n,n'}(l_{\rm B}^2q_{\|}^2/2)  
\Pi _2(n,n',\Omega),
\label{pi_2}\\
&&\hspace{-0.7cm}\Pi _2(n,n',\Omega)=-\frac2\pi \int d\varepsilon
\left [ f(\varepsilon )- f(\varepsilon +\Omega)\right ]\nonumber\\
&&\,\hspace{2cm}\times\,\,{\rm Im}G_n(\varepsilon +\Omega){\rm Im}G_{n'}(\varepsilon ),
\end{eqnarray}
where $l_{\rm B}=\sqrt{1/|eB|}$ is the magnetic length,
\begin{equation}
C_{n,n+l}(Y)\equiv n![(n+l)!]^{-1}Y^le^{-Y}[L_n^l(Y)]^2\label{cnn}
\end{equation}
with $L_n^l(Y)$ the associate Laguerre polynomial, $f(\varepsilon
)=\{\exp [(\varepsilon -\mu)/T_{\rm e}]+1\}^{-1}$ the Fermi distribution
function, and ${\rm Im}G_n(\varepsilon )$ is the imaginary part of the 
electron Green's function, or the DOS, of the Landau level $n$.
The real part function $\Pi_{01}({\bf q}_{\|},\Omega)$ and corresponding
$\Lambda_{01}({\bf q}_{\|},\Omega)$ function can be derived from their  
imaginary parts via the Kramers-Kronig relation.

In principle, to obtain the Green's function ${\rm Im}G_n(\varepsilon )$,
a self-consistent calculation has to be carried out from the Dyson equation 
for the self-energy with all the impurity, phonon and other scatterings 
included. The resultant $G_n$ is generally a complicated function of
the magnetic field, temperature, and Landau-level index $n$, also
dependent on the relative strengths of different kinds of 
scatterings.\cite{Ando82,Leadley}
In the present study we do not attempt a self-consistent calculation of
$G_n(\varepsilon)$. Instead, we choose a Gaussian-type form\cite{Ando82}
for the purpose of demonstrating the observed oscillations ($\varepsilon_n$
is the energy of the $n$-th Landau level):
\begin{equation}
{\rm Im}G_n(\varepsilon)=-\sqrt{\frac{\pi}{2\Gamma^2}}
\exp\Big[-\frac{(\varepsilon-\varepsilon_n)^2}{2\Gamma^2}\Big]
\end{equation}
with a broadening width giving by
\begin{equation}
\Gamma=\Big(\frac{2e\omega_c\alpha}{\pi m \mu_0(T)}\Big)^{1/2},\label{gamma}
\end{equation}
where $\mu_0(T)$ is the linear mobility at temperature $T$ 
in the absence of the magnetic field and $\alpha > 1$ is a semiempirical 
parameter to take account the difference 
of the transport scattering time determining the mobility $\mu_0(T)$, to which
larger angle scattering contributes a heavier weight,    
from the single particle lifetime, to which scattering with small or large angle
equally contributes.\cite{Mani,Durst,Anderson}.

\section{GaAs-based 2DES}

To contact with recent experiments, we focus our attention to two ultra 
high mobility two-dimensional electron systems of GaAs/AlGaAs heterostructure 
with same electron sheet density
$N_{\rm e}=3\times 10^{11}\,$cm$^{-2}$ but having linear mobility
$\mu_0 (1\,{\rm K})=2.4\times 10^{7}\,$cm$^2\,$V$^{-1}$s$^{-1}$ and 
$\mu_0 (1\,{\rm K})=1.46\times 10^{7}\,$cm$^2\,$V$^{-1}$s$^{-1}$ respectively 
in the absence of the magnetic field. 
In GaAs/AlGaAs systems, phonon modes and electron-phonon couplings are 
well established.\cite{Lei851} We consider both transverse acoustic 
phonons (interacting with electrons via piezoelectric coupling), and longitudinal
acoustic phonons (interacting with electrons via piezoelectric and 
deformation potential couplings). 
We assumed that the elastic scatterings are either due to 
the remote charged impurities which are locate a distance $s=60$\,nm away 
from the interface of the heterojunction in the barrier side, 
or due to background impurities which are evenly distributed throughout
the GaAs region. The impurity densities are determined by the 
requirement that electron total linear mobility equals the giving value at
temperature $T=1$\,K.  
The relevant effective impurity scattering potentials $|U({\bf q}_{\|})|^2$
and electron-phonon matrix elements $|M({\bf q})|^2$ were discussed 
earlier.\cite{Lei851} 
The material and coupling parameters for the system are well defined and 
taken as: electron effective mass $m=0.068\,m_{\rm e}$ ($m_{\rm e}$ is the
free electron mass), transverse sound speed $v_{\rm st}=2.48\times 10^3$\,m/s,
longitudinal sound speed $v_{\rm sl}=5.29\times 10^3$\,m/s, acoustic 
deformation potential $\Xi=8.5$\,eV, piezoelectric constant $e_{14}=
1.41\times 10^9$\,V/m, dielectric constant $\kappa=12.9$, 
material mass density $d=5.31$\,g/cm$^3$.
The depletion layer charge number density is taken as 
$N_{\rm dep}=5\times 10^{10}$\,cm$^{-2}$.

 In GaAs systems at low temperatures 
$\mu_0(T)$ comes from impurity, transverse and longitudinal acoustic phonon 
scatterings:
\begin{equation}
\frac{1}{\mu_0}= \frac{1}{\mu_0^{\rm (i)}}+\frac{1}{\mu_0^{\rm 
(pt)}}+\frac{1}{\mu_0^{\rm (pl)}}. 
\end{equation}
The longitudinal 
magnetoresistivity $R_{xx}$ can also be formally written as the sum of the impurity, 
transverse and longitudinal acoustic phonon contributions:
\begin{equation}
R_{xx}=R_{xx}^{\rm (i)}+R_{xx}^{\rm (pt)}+R_{xx}^{\rm (pl)}.
\end{equation} 

In the following sections we will carried out numerical calculations
for $R_{xx}^{\rm (i)}$, $R_{xx}^{\rm (pt)}$ and $R_{xx}^{\rm (pl)}$
assuming linearly polarized MW fields (${\bf E}_c=0$) 
with multiphoton processes included.

The microwave field intensity required for the appearance of resistivity 
oscillation in these high-mobility samples is moderate. 
The slight electron heating induced by the irradiation in these systems is
unimportant as far as the main phenomenon is concerned. As can be seen later 
that the radiation-induced magnetoresistance oscillation is not sensitive 
to the moderate rise of electron temperature. Therefore, $R_{xx}$ can   
be obtained directly from Eq.\,(\ref{lrxx}) with $T_{\rm e}=T$. We will 
check the effect of elevated electron temperature in the subsection IV-c.

\section{$R_{xx}$ at temperature $T=1$\,K}

\subsection{Impurity-induced magnetoresistivity}

\subsubsection{Linear resistivity}

We first assume that the elastic scatterings are due to ionized remote 
impurities.\cite{Lei851}
Fig.\,1a shows the impurity-induced longitudinal resistivity $R_{xx}^{\rm (i)}$ 
versus $\omega/\omega_c\equiv \gamma_c$ subjected to a microwave radiation of frequency 
$\omega/2\pi=0.1$\,THz at four values of amplitude: 
$E_s=20, 45, 65$ and 80\,V/cm. At temperature $T=1$\,K
SdH oscillations of period $\gamma_c=0.039$ 
show up strongly at high $\omega_c$ side, 
and then gradually decay away as $1/\omega_c$ increases. In addition to this
all four resistivity curves exhibit clear oscillation having the main oscillation 
period 
$\gamma_c=1$ (they are crossing at integer points $\gamma_c=2,3,4,5$).
The resistivity maxima locate around $\gamma_c=j-\delta_{-}$ 
and minima around $\gamma_c=j+\delta_{+}$ with $\delta_{\pm}\sim 0.22-0.26$
for $j=3,4,5$, $\delta_{\pm}\sim 0.17-0.22$ for $j=2$, 
and $\delta_{\pm}\sim 0.08-0.14$ for $j=1$ (Fig.\,1b). 
The amplitude of the oscillation increases with increasing
HF field intensity for $\gamma_c >1.5$. Resistivity gets into negative value for 
$E_s=80$\,V/cm around the minima at $j=1,2$ and 3, for $E_s=65$\,V/cm at $j=1$ and 2,
and for $E_s=20$ and 45\,V/cm at $j=1$.
These main peak-valley structures are related to single-photon ($|n|=1$) processes.
In the vicinity of $\gamma_c=1$, where the cyclotron resonance greatly 
enhances the effective amplitude of the HF field in photon-assisted 
scatterings, multiphoton processes show up.
The amplitudes of the $j=1$ maximum and minimum no longer monotonically change with
field intensity. Furthermore, there appears a shoulder around 
$\gamma_c=1.5$ on the curves of $E_s=45$ and 65\,V/cm, and it develops into
a secondary peak in the $E_s=80$\,V/cm case. This peak-valley structure around 
$\gamma_c=1.5$ is related to two-photon ($|n|= 2$) processes.
The oscillatory peak-valley structure related to three-photon processes
was also demonstrated in the cases of lower microwave frequency ($\omega=60$ and 
40\,GHz).\cite{Lei04687}
 
 The dependence of the resistivity at maxima and minima on the microwave intensity
 is shown in Fig.\,1c, where we plot the calculated photoresistivity, i.e. the
 magnetoresistivity in the presence of radiation, $R_{xx}^{\rm (i)}$, 
 minus the dark resistivity $R_{xx}^{\rm (i)}(E_s=0)$, as a function of radiation field 
 amplitude $E_{s}$ at peaks and valleys of $j=2$ and $j=3$. We see that 
$|R_{xx}-R_{xx}(0)|$
 grows like $E_s^2$ at lower intensity and become linearly dependent on $E_s$
 at higher intensity within the amplitude range shown. This is in agreement with
 experiments.\cite{Mani,Zud03,Mani06388} 
 
 Fig.\,1d shows the positive parts of $R_{xx}^{\rm (i)}/R_{xx}^{\rm (i)}(E_s=0)$ at 
minima of $j=2,3$ and 4 on logarithmic scale as functions of $E_s^2$. 
  
\begin{figure}
\includegraphics [width=0.45\textwidth,clip=on] {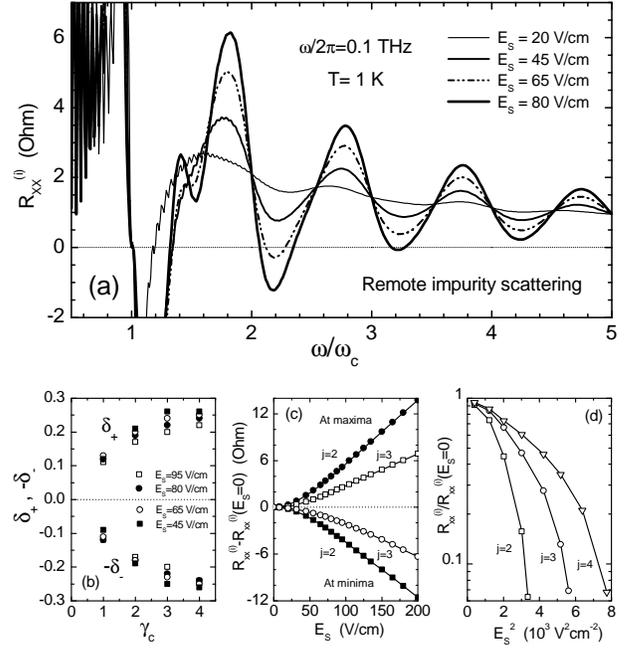}
\vspace*{-0.2cm}
\caption{(a) The longitudinal linear magnetoresistivity $R_{xx}^{\rm (i)}$ induced by 
remote-impurity scattering in a GaAs-based heterosystem  
subjected to crossed magnetic fields $B$ and in-plane linearly polarized HF 
fields $E_s\sin(\omega t)$ of frequency $\omega/2\pi=0.1$\,THz with several 
different amplitudes at lattice temperature $T=1$\,K. $\omega_c\equiv eB/m$ 
stands for the cyclotron frequency. The other parameters are:  
electron density $N_{\rm e}=3.0\times 10^{11}$\,cm$^{-2}$, zero-magnetic-field 
linear dc mobility $\mu_0(1\,{\rm K})=2.4\times 10^7$\,cm$^2$ V$^{-1}$ s$^{-1}$, 
and the broadening coefficient $\alpha=12$. The electron temperature is set to be 
$T_{\rm e}=T$.
(b) Parameters $\delta_{+}$ and $\delta_{-}$ for locations of resistance maxima and
 minima at several HF field amplitudes. (c) The photoresistivity 
 $R_{xx}^{\rm (i)}-R_{xx}^{\rm (i)}(E_s=0)$ at maxima and at minima of 
 $j=1$ and $j=2$ against the amplitude of the HF field. 
 (d) $R_{xx}^{\rm (i)}/R_{xx}^{\rm (i)}(E_s=0)$ is shown against
$E_s^2$ on logarithmic scale for $j=2,3$ and 4.}
\label{fig1}
\end{figure}

\begin{figure}
\includegraphics [width=0.48\textwidth,clip=on] {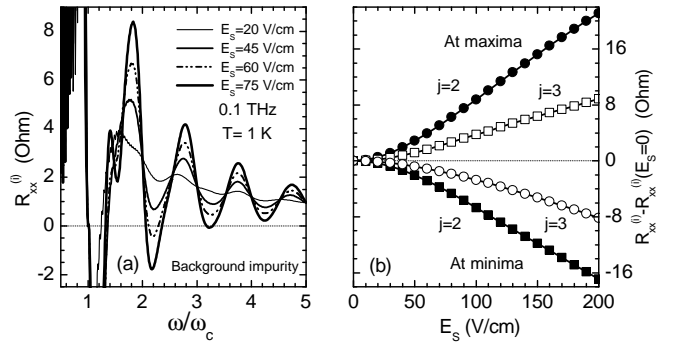}
\vspace*{-0.2cm}
\caption{(a) The linear magnetoresistivity $R_{xx}^{\rm (i)}$ induced by 
background-impurity scattering in a GaAs-based 2DEG subjected to linearly 
polarized HF fields $E_s\sin(\omega t)$ of frequency $\omega/2\pi=0.1$\,THz. 
The system parameters are the same as indicated in Fig.\,1. 
(b) The background impurity-induced photoresistivity 
$R_{xx}^{\rm (i)}-R_{xx}^{\rm (i)}(0)$ at maxima and at minima of $j=1$ and $j=2$ 
shown in Fig.\,2a is plotted against the amplitude of the HF field.}
\label{fig2}
\end{figure}

 It should be noted that the above feature of impurity-induced magnetoresistivity
 is not very sensitive to the form of elastic scattering potential. In Fig.\,2a
 we demonstrate $R_{xx}^{\rm (i)}$ due to background impurity scattering as a function
 of $\omega/\omega_c$ under several different microwave amplitudes. Although
 the scattering potential of background impurities appears quite different from that  
 of remote impurities,\cite{Lei851} the main features of the radiation-induced 
 magnetoresistivity oscillation are in common for both scattering potentials. 
 The difference shows up only in some quantitative details. 
 The dependence of the background impurity induced resistivity 
 at maxima and minima on the microwave intensity
 shown in Fig.\,2b, is also similar to the case of remote impurity scattering
 (Fig.\,1c).

 \subsubsection{Nonlinear resistivity} 
 
 \begin{figure}
\includegraphics [width=0.45\textwidth,clip=on] {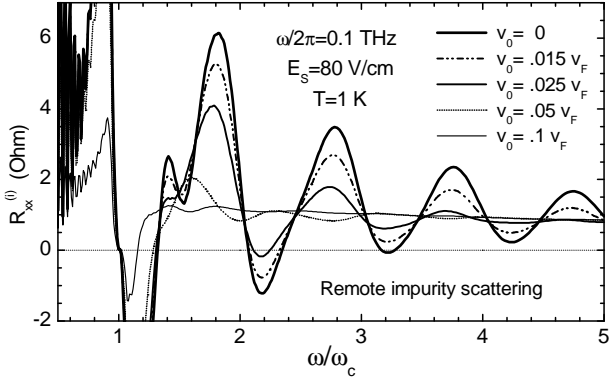}
\vspace*{-0.2cm}
\caption{Remote-impurity induced nonlinear longitudinal magnetoresistivity 
$R_{xx}^{\rm (i)}$ as defined in Eq.(\ref{rxx}) in the GaAs-based 2DEG   
subjected to crossed magnetic fields $B$ and an in-plane linearly polarized 
HF field $E_s\sin(\omega t)$ of frequency $\omega/2\pi=0.1$\,THz and amplitude 
$E_s=80$\,V/cm
under several different dc bias velocities $v_0=0,.015,.025,.05$ and .1\,$v_F$,
where $v_F=2.4\times 10^5$\,m/s is the electron Fermi velocity. 
$\omega_c\equiv eB/m$ is the cyclotron frequency.
The other parameters are the same as indicated in Fig.\,1.}
\label{fig3}
\end{figure}
 
  Fig.\,3 shows the nonlinear longitudinal resistivity 
  $R_{xx}^{\rm (i)}$ due to remote-impurity scattering calculated directly from 
Eq.\,(\ref{rxx}) for the 2DES subject to a 0.1\,THz microwave radiation 
 of amplitude $E_s=80$\,V/cm under different dc bias velocities 
 $v_0=0, .015,.025,.05$ and 0.1\,$v_{\rm F}$, 
 where the electron Fermi velocity $v_{\rm F}=2.4\times 10^5$\,m/s. 
For the given strength of the radiation field the linear magnetoresistivity (vanishing
$E_0$ or $v_0$) exhibits the strongest oscillation. A finite dc bias always 
suppresses the oscillation and may destroy the negative resistivity appearing 
at vanishing dc bias.

\subsection{Acoustic-phonon-induced magnetoresistivity} 

Acoustic phonon scattering has recently been proposed as a mechanism of 
the absolute negative resistivity leading to vanishing resistance state.\cite{Ryz05199,
Ryz05454} To check such a possibility  
we show in Fig.\,4 the linear magnetoresistivity $R_{xx}$, 
contributed separately by transverse acoustic phonon scattering (a) and 
by longitudinal acoustic phonon scattering (b), as functions of $\omega/\omega_c$ 
under 0.1\,THz microwave illumination of different strengths.
Photon-assisted acoustic-phonon scattering itself indeed can give rise to pronounced  
resistance oscillation with changing magnetic field and $R_{xx}$ at oscillation minima 
can go down to negative under sufficiently strong microwave irradiation 
for both transverse and longitudinal phonon scatterings. These acoustic-phonon 
induced $R_{xx}$ oscillations, 
though can still be considered nearly periodical in inverse magnetic field,
exhibit quite different behavior from that of impurity-induced $R_{xx}$ oscillation
shown in Fig.\,1 and also different from each other, especially in respect to 
the phase of the oscillation. In the case of impurity scattering 
the resistance maxima always appear at $\gamma_c=j-\delta_{-}$ and 
the maxima at $\gamma_c=j+\delta_{+}$,
while in the case of longitudinal acoustic scattering the  maxima
appear at $\gamma_c=j+\delta_{+}$ and the minima at $\gamma_c=j-\delta_{-}$
except the case of $j=1$. The phase of transverse acoustic phonon induced 
$R_{xx}$ oscillation resembles that of impurity at $j=1,2$ but becomes
that of longitudinal acoustic phonon at $j=3,4,5$.

\begin{figure}
\includegraphics [width=0.4\textwidth,clip=on] {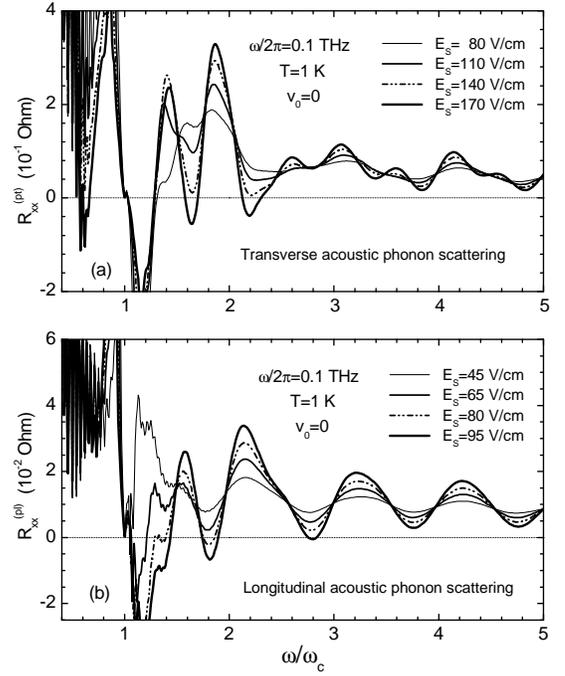}
\vspace*{-0.2cm}
\caption{Linear magnetoresistivities 
$R_{xx}^{\rm (pt)}$ (a) and $R_{xx}^{\rm (pl)}$ (b) induced by transverse
and longitudinal acoustic phonons in a GaAs-based 2DES subjected to 
microwave fields $E_s\sin(\omega t)$ of frequency $\omega/2\pi= 0.1$\,THz
having different amplitudes. The lattice temperature is $T=1$\,K and  
the electron temperature $T_{\rm e}=T$. The other parameters are:  
electron density $N_{\rm e}=3.0\times 10^{11}$\,cm$^{-2}$, dc mobility
$\mu_0(1\,{\rm K})=2.4\times 10^7$\,cm$^2$ V$^{-1}$ s$^{-1}$, and the broadening 
coefficient $\alpha=12$.}
\label{fig4}
\end{figure}

As in the case of impurity scattering a finite dc bias always 
suppresses the phonon-induced resistance oscillations and 
may destroy the negative resistivity appearing at vanishing dc bias 
as seen in Fig.\,5, where we plot
the nonlinear longitudinal resistivity $R_{xx}^{\rm (pt)}$ induced by transverse
acoustic phonons and $R_{xx}^{\rm (pl)}$ induced by longitudinal acoustic phonons,
for the 2DEG subject to a 0.1\,THz microwave radiation with fixed amplitudes 
under several different dc current biases. These results are in qualitative 
agreement with a recent report.\cite{Ryz06051} 

\begin{figure}
\includegraphics [width=0.4\textwidth,clip=on] {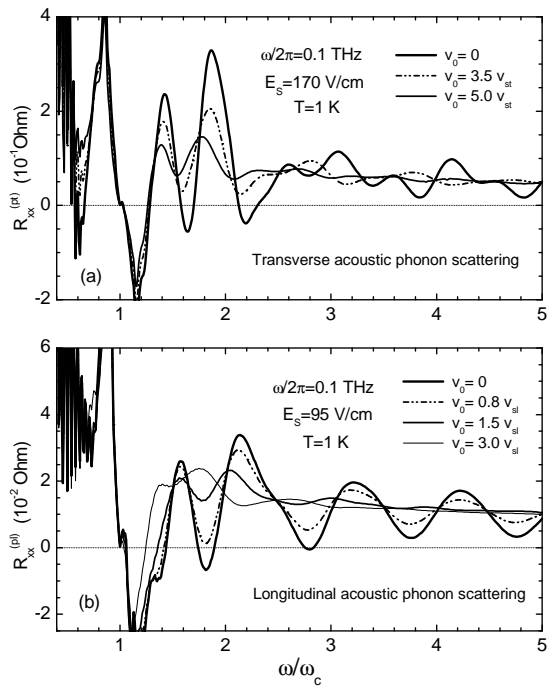}
\vspace*{-0.2cm}
\caption{ Nonlinear magnetoresistivities  
$R_{xx}^{\rm (pt)}$ (a) and $R_{xx}^{\rm (pl)}$ (b) induced 
by transverse and longitudinal acoustic phonons under different  
dc bias $v_0$ in a GaAs-based 2DES exposed to microwave fields 
$E_s\sin(\omega t)$ of frequency $\omega/2\pi= 0.1$\,THz 
having amplitude $E_s=170$\,V/cm and $E_s=95$\,V/cm respectively.
Here $v_{\rm st}= 2.48\times 10^3$\,m/s and 
$v_{\rm sl}=5.29\times 10^3$\,m/s are the transverse and longitudinal 
sound speed. The other parameters are
the same as indicated in Fig.\,4.}
\label{fig5}
\end{figure}

However, at temperature $T=1$\,K the acoustic phonon scattering
contributes a part of $R_{xx}$ which is more than an order of magnitude smaller
than that contributed from impurity scattering in the system having a mobility
of $2.4\times 10^7$\,cm$^2$V$^{-1}$s$^{-1}$ at $T=1$\,K. 
Therefore, acoustic phonon scattering
essentially gives no direct contribution to the experimentally observed resistance
oscillation at this low temperature. Nevertheless, acoustic phonons play a key role 
in suppressing the resistance oscillation at elevated lattice temperatures. 
We will discuss this issue in Section V.

\subsection{Effect of elevated electron temperature}                           
 
\begin{figure}
\includegraphics [width=0.48\textwidth,clip=on] {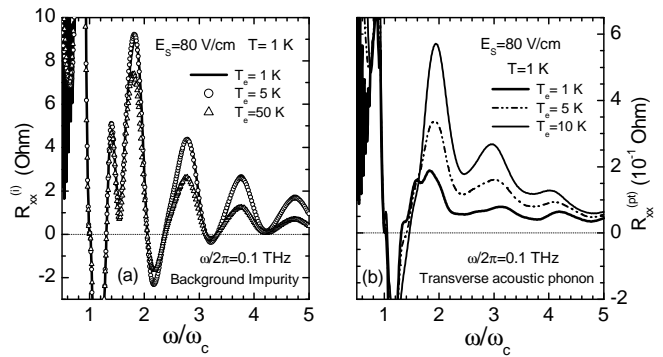}
\vspace*{-0.2cm}
\caption{The impurity-induced linear magnetoresistivity 
$R_{xx}^{\rm (i)}$ (a) and transverse acoustic phonon induced linear 
magnetoresistivity $R_{xx}^{\rm (pt)}$
of a GaAs-based 2DES subjected to a microwave field of amplitude 
$E_s=80$\,V/cm and frequency $\omega/2\pi= 0.1$\,THz
at elevated electron temperatures.
The lattice temperature is $T=1$\,K. The other parameters are
the same as indicated in Fig.\,4.}
\label{fig6}
\end{figure}
 
 Upon the microwave irradiation the electron temperature can be higher than
 the lattice temperature. To have an idea of how the elevated electron 
 temperature affects the magnetoresistivity oscillation we show in Fig.\,6a
 the background impurity induced longitudinal resisitivity $R_{xx}^{\rm (i)}$ 
 of the 2DEG having several different electron temperatures $T_{\rm e}=1,5$ and 50\,K
 but all at the same lattice temperature $T=1$\,K  subject to a microwave irradiation 
 of frequency $\omega/2\pi=0.1$\,THz and amplitude $E_s=80$\,kV/cm. 
 We see that at $T_{\rm e}=5$\,K (even much lower) SdH oscillations disappear
completely, but the radiation-induced oscillations remain essentially the same.
Only when the electron temperature becomes much higher, e.g. 50\,K, can appear 
the appreciable change in the resistance oscillation curve. This indicates that 
the radiation-induced resistivity oscillation is quite insensitive to electron 
temperature and we can analyze the moderate-strength microwave induced 
resistivity oscillation by neglecting the electron temperature change
in the system. On the other hand, using a slightly elevated electron temperature  
provides a way to separate the radiation-induced oscillation from the SdH effect.

Fig.\,6b shows the effect of elevated electron temperature on transverse acoustic 
phonon induced resistivity $R_{xx}^{\rm (pt)}$. With electron temperature increasing
from $T_{\rm e}=1$\,K to 5\,K, the magnitude of the oscillation somewhat grows.
But its absolute contribution to magnetoresistivity is still an order of magnitude 
smaller than the of impurity as long as the lattice temperature remains $T=1$\,K. 
 
Experiments showed that the strongest SdH oscillations always show up 
on the dark resistivity curve and with enhancing the radiation intensity 
the SdH oscillations weaken.\cite{Mani,Mani06388}
This fact can be easily understood as due to the rise of the 
electron temperature caused by the microwave illumination: SdH oscillations
are suppressed by the rising electron temperature, while the radiation-induced
$R_{xx}$ oscillations remain the same as long as the lattice temperature keeps
unchange.

\section{Effect of elevated lattice temperature}

One of the most important aspect of the experimental finding on the magnetoresistance
oscillation in irradiated 2DES is the temperature-dependence of $R_{xx}$.
Experiments\cite{Mani,Zud03} found that 
the "zero-resistance" states and radiation-induced magnetoresistance oscillations 
show up strongly only at low temperatures typically around $T=1$\,K or lower. 
At fixed microwave power with increasing temperature, not only the zero-resistance 
regions become narrower and eventually disappear, the whole oscillatory 
structure (peaks and valleys) diminish as well. At temperature $T\geq$\,4-5\,K,
oscillatory structure disappears completely and the resistivivty $R_{xx}$-versus
magnetic field becomes essentially flat.\cite{Zud03} 

Both groups analyzed the temperature-variation of $R_{xx}$ at deepest minima 
using an activated-type dependence $\exp(T_0/T)$. The activation energies
observed by both groups are very high and different: up to 10\,K and 20\,K 
at $j=1$ minimum respectively.\cite{Mani,Zud03} These values are
about an order of magnitude higher than the microwave photon energy 
($\omega\sim $3-5\,K) and Landau-level spacing ($\omega_c\leq$2\,K). 
The different $T_0$ values observed by the two groups indicate that the 
speed of the oscillatory structure disappearing with temperature 
is sample dependent.\cite{Mani,Zud03}
To explain the temperature dependence the formation of an energy gap around
the Fermi surface in the spectrum is suggested under microwave irradiation 
around the resistance minima.\cite{Mani}

Our explanation of the temperature dependence of the magnetoresistance oscillations
is based on the temperature variation of the Landau level broadening $\Gamma$
as determined by Eq.\,(\ref{gamma}). In a GaAs-based system, when the lattice 
temperature 
increases from around $T=1$\,K, the numbers of transverse and longitudinal 
acoustic phonons and thus the electron-phonon scattering strengths increase
rapidly. In Fig.\,7b we plot the zero-magnetic-field linear
mobility $\mu_0^{\rm (pt)}$ due to transverse acoustic phonon scattering,
$\mu_0^{\rm (pl)}$ due to longitudinal phonon scattering, and the total 
mobility $\mu_0$ as functions of lattice temperature $T$ for the GaAs-based 
heterosystem 
with $\mu_0=2.4\times 10^7$\,cm$^2$V$^{-1}$s$^{-1}$ at $T=1$\,K. We see that 
when temperature $T$ rises from 1\,K to 3\,K the phonon related mobilities decline  
about an order of magnitude, leading to
the total mobility $\mu_0(T)$ decreasing about a factor of 2.2 and,
according to (\ref{gamma}), $\Gamma$ increasing about a factor of 1.5
(assuming $\alpha$ unchanged).   
The temperature growth of the Landau-level width due to this 
enhanced phonon scattering results in the strong temperature variation of
the radiation-induced magnetoresistance oscillation.
The effect of such a mobility decrease or the Landau level broadening on
$R_{xx}$ is significant. Fig.\,7a shows the calculated linear resistivity 
$R_{xx}$ as a function of $\omega/\omega_c$ at different lattice temperatures
 $T=1.0,1.5,2.0,2.5,3.0,3.5,4.0$ and 4.5\,K for the system of 
$\mu_0(1\,{\rm K})=2.4\times 10^7$cm$^2$V$^{-1}$s$^{-1}$ 
under a fixed microwave illumination of
frequency $\omega/2\pi=0.1$\,THz and amplitude $E_s=80$\,V/cm. The broadening
parameter is fixed to be $\alpha=12$ for all the curves of different lattice 
temperatures, and the electron temperature is taken to equal to the lattice 
temperature in all the calculations in this section. The sensitive temperature
dependence of the resistance oscillation is quite obvious. The magnitude of 
peaks and valleys of the oscillation diminish straightforward with increasing 
temperature from 1\,K. At $T \ge 4.5$\,K the oscillation structure almost 
disappear and $R_{xx}$ exhibits quite a flat form with changing 
$\omega/\omega_c$ for $\gamma_c \ge 1.6$.

\begin{figure}
\includegraphics [width=0.45\textwidth,clip=on] {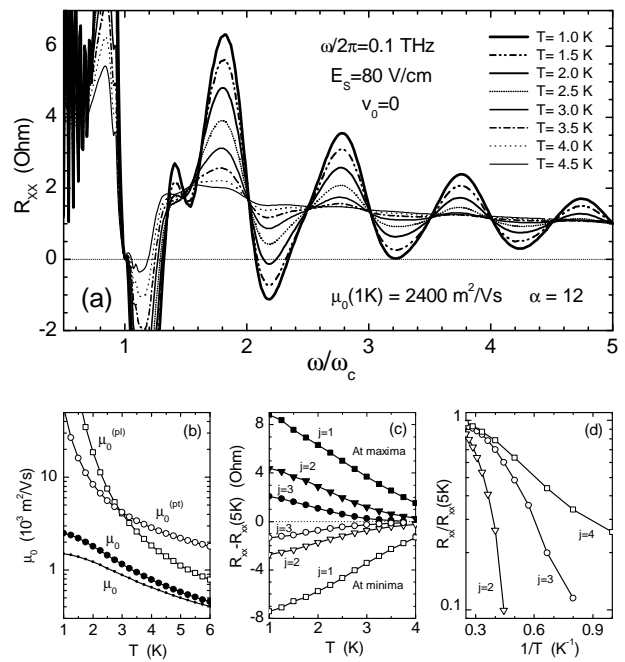}
\vspace*{-0.2cm}
\caption{(a) The longitudinal magnetoresistivity $R_{xx}$ at different lattice 
temperatures $T=1.0,1.5,2.0,2.5,3.0,3.5,4.0$ and 4.5\,K for a GaAs-based 2DEG 
subjected to a HF field $E_s\sin(\omega t)$ of frequency $\omega/2\pi=0.1$\,THz
and amplitude $E_s=80$\,V/cm. The system parameters are:  
electron density $N_{\rm e}=3.0\times 10^{11}$\,cm$^{-2}$, dc mobility
$\mu_0=2.4\times 10^7$\,cm$^2$ V$^{-1}$ s$^{-1}$ at $T=1$\,K, and the broadening 
coefficient $\alpha=12$ for all the curves. (b) Zero magnetic field linear mobility
induced by transverse acoustic phonon scattering, $\mu_0^{\rm (pt)}$, by longitudinal
acoustic phonon scattering, $\mu_0^{\rm (pl)}$, and the total mobility $\mu_0$, 
for the system with $\mu_0(1\,{\rm K})=2.4\times 10^7$cm$^2$V$^{-1}$ s$^{-1}$ (dots), 
and for system 
with $\mu_0(1\,{\rm K})=1.46\times 10^7$cm$^2$V$^{-1}$ s$^{-1}$ (solid line). (c) The 
amplitudes of the 
resistivity oscillation $R_{xx}(T)-R_{xx}(5\,{\rm K})$ at maxima and at minima versus 
temperature 
$T$ for $j=1,2$ and 3. (d) $R_{xx}(T)/R_{xx}(5\,{\rm K})$ is shown against
$1/T$ on logarithmic scale for $j=2,3$ and 4.}
\label{fig7}
\end{figure}
\begin{figure}
\includegraphics [width=0.45\textwidth,clip=on] {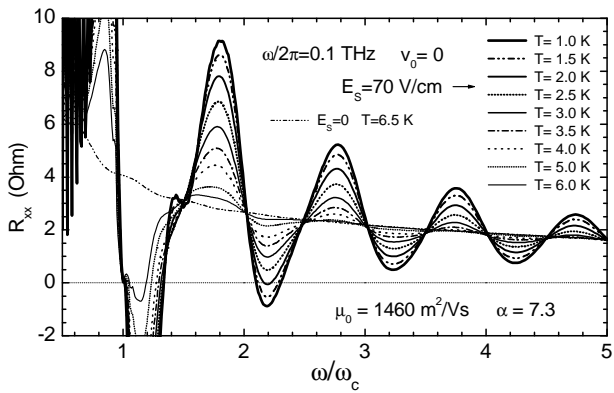}
\vspace*{-0.2cm}
\caption{The longitudinal magnetoresistivity $R_{xx}$ at different lattice temperature 
$T=1.0,1.5,2.0,2.5,3.0,3.5,4.0,5.0$ and 6.0\,K for a GaAs-based 2DEG 
subjected to a HF field $E_s\sin(\omega t)$ of frequency $\omega/2\pi=0.1$\,THz
and amplitude $E_s=70$\,V/cm. The system parameters are:  
electron density $N_{\rm e}=3.0\times 10^{11}$\,cm$^{-2}$, dc mobility
$\mu_0(1\,{\rm K})=1.46\times 10^7$\,cm$^2$ V$^{-1}$ s$^{-1}$, and the 
broadening coefficient 
$\alpha=7.3$ for all the curves. Also shown is the dark resistivity
($E_s=0$) at $T=6.5$\,K.}
\label{fig8}
\end{figure}

Comparing Fig.\,7a with Fig.\,1a,\cite{exp} we see that the role of decreasing 
inverse temperature at fixed radiation intensity is somewhat similar to that 
of reducing microwave intensity at fixed temperature. This similarity, however,
holds only within the range of $T\leq$4-5\,K when direct phonon contribution 
to $R_{xx}$ is still less important than that of impurities. At higher temperature 
the direct phonon contribution may become dominant and the structureless $R_{xx}$
will continue to grow with rising $T$ and the $R_{xx}$-vs-$\gamma_c$ curves of
different $T$ will not cross at integer points $\gamma_c=2,3,4,5$, a feature
of impurity scattering dominance.     

To show the temperature variation of $R_{xx}$, we plot in Fig.\,7c 
$R_{xx}-R_{xx}(5\,{\rm K})$ at maxima and at minima of $j=1,2$ and 3, as function of 
$T$.
The oscillation disappears faster at larger $j$ (lower magnetic field) than at
smaller $j$ (higher magnetic field). The positive values of $R_{xx}/R_{xx}(5\,{\rm K})$ 
at peaks and at valleys also plotted against $1/T$ on logarithmic scale. If we roughly
fit the data with the form $R_{xx}(T)\propto \exp(-T_0/T)$, we have $T_0\approx 10$\,K
for $j=2$,  $T_0\approx 4.5$\,K for $j=3$, and $T_0\approx 2.1$\,K 
for $j=4$ in average over the range shown. 

The sensitivity of the temperature variation of the radiation-induced magnetoresistance
oscillation is sample dependent. In GaAs-based systems the electron-phonon scattering 
strengths, thus the acoustic-phonon induced mobilities $\mu_0^{\rm (pt)}$ and 
$\mu_0^{\rm (pl)}$ 
and their temperature behavior are essentially the same. Therefore, the temperature 
variation of the total mobility $\mu_0$ depends mainly on the strength of the impurity
scattering, which is almost temperature independent within this range of $T$.
The lowest curve in fig.\,7b shows the $T$-dependence of the total mobility $\mu_0(T)$ 
for the sample having $T=1$\,K mobility $\mu_0(1\,{\rm K})=1.46\times 
10^{7}$cm$^{2}$V$^{-1}$s$^{-1}$, 
which apparently exhibits a slower temperature change than that of 
$\mu_0(1\,{\rm K})=2.4\times 10^{7}$cm$^{2}$V$^{-1}$s$^{-1}$ system. 

In Fig.\,8 we illustrate the linear resistivity 
$R_{xx}$ as a function of $\omega/\omega_c$ at different lattice temperatures
 $T=1.0,1.5,2.0,2.5,3.0,3.5,4.0,5.0$ and 6.0\,K for the system of 
$\mu_0(1\,{\rm K})=1.46\times 10^7$cm$^2$V$^{-1}$s$^{-1}$ under a fixed microwave 
irradiation of
frequency $\omega/2\pi=0.1$\,THz and amplitude $E_s=70$\,V/cm. The broadening
parameter is fixed to be $\alpha=7.3$ for all the lattice temperatures.
The speed of the oscillatory structure disappearing with rising temperature is 
apparently
slower than the system shown in Fig.\,7a. We see that at $T \ge 6.0$\,K the oscillation 
structure essentially disappear and $R_{xx}$ approaches a flat form of 
dark ($E_s=0$) curve of corresponding temperature (see Fig.\,8) for $\gamma_c \ge 1.6$. 
Note that different from $T=1$\,K case, where impurity scattering dominates, 
at $T=6$\,K the acoustic phonon
scatterings already yield a substantial contribution to both zero-field mobility 
$\mu_0$ and to
radiation-induced magnetoresistivity $R_{xx}$.

\section{Conclusion}

Based on the balance-equation model for magnetotransport
in Faraday geometry, we have carried out a detailed theoretical investigation on 
microwave-radiation induced magnetoresistance oscillations recently discovered
in high-mobility GaAs-based two-dimensional electron systems.
We find that for systems having zero-field linear mobility 
$\mu_0 (1\,{\rm K})\le 2.4\times 10^7$cm$^2$V$^{-1}$s$^{-1}$  
multiphoton-assisted impurity scatterings are the main mechanisms responsible for 
radiation-induced magnetoresistance oscillations at temperature $T\leq$4\,K. 
The amplitude of the $R_{xx}$ oscillation grows roughly
following the microwave power under weak illumination and following 
the microwave amplitude 
under medium illumination before it saturates even decreases with 
continuing increase of the microwave strength under strong irradiation. 
It is shown that the strongest oscillations appear in the linear longitudinal
 magnetoresistance and a finite dc current bias always suppresses the oscillation. 
We find that, different from the SdH oscillation which is easily suppressed by 
a few-degree rise of the electron temperature, the radiation-induced 
magnetoresistance oscillations are quite insensitive to the modest 
electron heating as long as the lattice temperature remain the same. 
Although the magnetoresistivities directly stemming from photon-assisted transverse 
and longitudinal acoustic phonon scatterings also exhibit pronounced oscillations 
under microwave irradiation, they contribute only a small part of the total $R_{xx}$
in the temperature range of $T\leq 4$\,K. 
Nevertheless, as we proposed, that it is just this acoustic phonon scattering that 
gives rise to the sensitive lattice temperature dependence of
 radiation-induced resistance oscillations right from $T=1$\,K.
We showed that the growth of the Landau level broadening resulting from the 
enhancement of acoustic phonon scatterings with increasing lattice temperature
leads to the observed temperature suppression of the oscillation.

\section*{Acknowledgments}

The author is grateful to Dr. S.Y. Liu for helpful discussions, to Prof. V.I. Ryzhii
for sending the information of Refs.\,\onlinecite{Ryz,Ryz86,Malov1973,Vyu1977}. 
This work was supported by the National Science Foundation of China,
the Special Funds for Major State Basic Research Project, and
the Shanghai Municipal Commission of Science and Technology.


\begin{thebibliography}{99}
\bibitem{exp} Since at lattice temperature $T=1$\,K, the phonon contribution 
to magnetoresistivity is almost negligible, the impuirity-induced resistivity 
$R_{xx}^{\rm (i)}$ shown in Fig.\,1 has only a minor difference from the total 
resistivity $R_{xx}$.
\end{thebibliography}
\end{document}